\documentclass[preprintnumbers,amsmath,amssymb,12pt,floatfix,epsfig,A4,nofootinbib]{revtex4}
\usepackage{bm}
\usepackage{natbib}
\usepackage[dvips]{color}
\usepackage{url}
\usepackage{graphicx}
\usepackage{amsmath}
\usepackage{float}
\usepackage{xcolor}
\usepackage{ulem}
\date{\today}

\usepackage{color} 

\newcommand\redout{\bgroup\markoverwith {\textcolor{red}{\rule[0.5ex]{2pt}{0.8pt}}}\ULon}

\begin{document}

\title{Suppression of the elastic scattering cross section for the $^{17}$Ne + $^{208}$Pb system}

\author{Kyoungsu Heo and Myung-Ki Cheoun}
\address{Department of Physics and Origin of Matter and Evolution of Galaxies (OMEG) Institute, Soongsil University, Seoul 06978, Korea}
\author{Ki-Seok Choi and K. S. Kim}
\address{School of Liberal Arts and Science, Korea Aerospace University, Koyang 10540, Korea}
\author{W. Y. So}
\thanks{\textrm{e-mail:} wyso@kangwon.ac.kr}
\address{Department of Radiological Science, Kangwon National University at Dogye, Samcheok 25945, Korea}
\date{\today}

\begin{abstract}
We investigated the elastic scattering, inelastic scattering, breakup reactions, and total fusion reactions of the $^{17}$Ne + $^{208}$Pb system using the extended optical model (OM) and a coupled channel (CC) approach. The aim of this study is to elucidate the suppression of the elastic cross section that is invisible in proton-rich nuclei, such as $^{8}$B and $^{17}$F projectiles, but appears in neutron-rich nuclei, such as $^{11}$Li and $^{11}$Be projectiles. The simultaneous $\chi^{2}$ analysis of the $^{17}$Ne + $^{208}$Pb system revealed that this suppression was primarily caused by the nuclear interaction between the projectile and the target nucleus, rather than the strong Coulomb interaction observed in neutron-rich nuclei. Moreover, the contributions of the Coulomb excitation interaction due to the two low-lying E2 resonance states were relatively small. In addition, the contribution of the direct reaction, comprising the inelastic scattering and breakup reaction cross sections, accounted for almost half of the total reaction cross-section. Finally, we performed the CC calculation using the parameters obtained from the OM calculation; however, the present CC calculations could not properly explain the $^{15}$O production cross section.

\end{abstract}

\pacs{24.10.-i, 25.70.Jj}
\maketitle

\section{Introduction}
\label{intro}
Owing to the recent development of the rare isotope (RI) technology, neutron- or proton-rich nuclei can be used as projectiles for collisions with heavy target nuclei, such as $^{6}$He + $^{208}$Pb~\cite{kaku03,sanc08}, $^{11}$Li + $^{208}$Pb~\cite{Cube12}, $^{11}$Be + $^{197}$Au~\cite{pesu17}, $^{8}$B + $^{\textrm{nat}}$Pb~\cite{yang13}, and $^{17}$F + $^{208}$Pb~\cite{Romo04, Lian02, Lian03} systems. A characteristic of these projectiles is their weak binding, which is reflected in their low separation energies.
Specifically, the separation energies of $^{6}$He ($S_{\textrm{2n}}$ = 0.975 MeV~\cite{kaku03}), $^{11}$Li ($S_{\textrm{2n}}$ = 0.369 MeV~\cite{Cube12}), $^{11}$Be ($S_{\textrm{n}}$ = 0.503 MeV~\cite{tani88}), $^{8}$B ($S_{\textrm{p}}$ = 0.138 MeV~\cite{yang13}), and $^{17}$F ($S_{\textrm{p}}$ = 0.601 MeV~\cite{Romo04}) are less than 1 MeV, which is considerably smaller than those of the well-known strongly bound $^{12}$C ($S_{\textrm{p}}$ = 15.956 MeV~\cite{wang17}) and $^{16}$O ($S_{\textrm{p}}$ = 12.127 MeV~\cite{wang17}) nuclei.

This implies that the projectile can be easily separated into core nuclei and valence nucleons (s) via Coulomb and/or nuclear interactions with the target nuclei. Consequently, the flux toward the elastic scattering channel is significantly diminished because of this strong breakup effect. This phenomenon is particularly prominent
in $^{11}$Li and $^{11}$Be projectiles with valence neutron(s), as shown in Fig.~3 of Ref.~\cite{Cube12} and Fig.~1 of Ref.~\cite{piet10}.
However, this phenomenon is not observed clearly for the $^{8}$B and $^{17}$F projectiles with valence protons. As shown in Fig. ~6 of Ref.~\cite{yang13}, the quasielastic scattering data of the $^{8}$B + $^{\textrm{nat}}$Pb system are significantly similar to those of the $^{7}$Be + $^{\textrm{nat}}$Pb system at $E_{\textrm{lab}}$/$V_{B}$ $\approx$ 3. Here, $E_{\textrm{lab}}$/$V_{B}$ denotes the ratio between the incident beam energy ($E_{\textrm{lab}}$) and the Coulomb barrier energy ($V_{B}$). This indicates that the breakup effect on elastic scattering is not substantial in the nuclear reaction of the proton-rich nuclei, as reported in Ref.~\cite{yang13}.

This raises the question of why the suppression of the elastic scattering cross sections fails to occur in the collision process between the heavy targets and proton-rich projectiles, such as $^{8}$B and $^{17}$F nuclei. One hypothesis for interpreting this phenomenon is the shielding effect~\cite{lian00,he20}. Fundamentally, the valence protons in $^{8}$B and $^{17}$F projectiles are repelled by the target nuclei. However, unlike valence neutrons, valence protons are strongly pushed back by Coulomb repulsion from the target nucleus, causing them to effectively hide behind the heavier core nucleus. This results in a shielding effect, in which the influence of the target nucleus is diminished. Consequently, the breakup reaction cross sections do not constitute a large proportion, as listed in Table II of Ref.~\cite{heo22}. This is in contrast to the phenomenon observed in $^{11}$Li and $^{11}$Be projectiles, in which the shielding effect is absent because of the presence of valence neutrons~\cite{so16,fern15}.

Recently, the elastic scattering data for the $^{17}$Ne + $^{\textrm{208}}$Pb system have been reported at energies close to the Coulomb barrier ($E_{\textrm{lab}}$/$V_{B}$ $\approx$ 1.17)~\cite{ovej23}, where the suppression effect in elastic scattering cross sections has been found in nuclear reactions with proton-rich nuclei. $^{17}$Ne is a proton-rich nucleus with a Borromean structure interconnected by three nuclei: $^{15}$O + p + p~\cite{ovej23,chro02,marg16}. In contrast to $^{8}$B and $^{17}$F nuclei, it can be inferred that the inelastic and/or breakup reactions significantly contribute toward the total reaction at energies close to the Coulomb barrier. This differs from the characteristics of nuclear reactions involving proton-rich nuclei reported thus far. Therefore, this study aims to examine the cause of the suppression of the elastic scattering cross sections of the proton-rich projectile $^{17}$Ne and $^{\textrm{208}}$Pb target nuclei system.

In Ref.~\cite{ovej23}, the elastic scattering data for the $^{17}$Ne + $^{\textrm{208}}$Pb system using the OM and CC methods, similar to our approach, were analyzed. However, the OM method used in the study involved fitting only the elastic scattering data, except for the breakup cross section, by employing a single real-volume components and a pair of imaginary volume and surface components. Consequently, although their outcomes could effectively describe the elastic scattering data, the reproduction of the breakup cross section was not achieved. For reference, we considered the $^{15}$O production cross section as the breakup cross section in our calculations. In addition, they attempted to reproduce the elastic scattering and breakup cross section by performing CC calculations, using the $\frac{3}{2}^{-}$ with $E_{x}$ = 1.288 MeV and $\frac{5}{2}^{-}$ with $E_{x}$ = 1.764 MeV resonance states.

Hence, in the present work, we aim to explain the elastic scattering and breakup cross-sectional data simultaneously using both the OM and CC methods, which are different from Ref.~\cite{ovej23}. In particular, the channels to be considered in the OM method are separated using the available experimental data. Using the separated channels, we calculate the elastic scattering, inelastic scattering, breakup reaction, and fusion cross sections, which were not shown in Ref.~\cite{ovej23}, and examine the contribution of these channels. Furthermore, the elastic scattering and breakup cross-section data are explained in the CC calculation by considering the $\frac{5}{2}^{-}$ state with $E_{x}$ = 2.692 MeV and the $\frac{3}{2}^{-}$ and $\frac{5}{2}^{-}$ resonance states employed in Ref.~\cite{ovej23}. For reference, there are two excited states ($\frac{1}{2}^{+}$ and $\frac{3}{2}^{+}$) associated with E1 transition~\cite{marg16}. However, since its $B(E1)$ values are very small [$B(E1)$ $<$ 0.007 $\textrm{e}^{2}\textrm{fm}^{2}$ for $\frac{1}{2}^{+}$ state and $B(E1)$ = 0.071 $\textrm{e}^{2}\textrm{fm}^{2}$ for $\frac{3}{2}^{+}$ state], these two excited states will not be considered in the present work.

The remainder of this paper is organized as follows. In Sec.~\ref{sec3}, we explain the formalism of the optical model (OM) used in this study. Based on all available experimental data, we evaluate the various cross sections related to elastic scattering, inelastic scattering, breakup reactions, and fusion reactions from the OM. In Sec.~\ref{simple approach}, we introduce dynamic polarization potentials (DPPs) to explain the reaction channels. Next, we discuss the characteristics of the proton-rich projectile, $^{17}$Ne, obtained by OM and CC frameworks. Finally, the conclusions of the study are presented in Sec.~\ref{con}.

\section{Formalism}
\label{sec3}
To investigate the suppression of the experimental $P_{\textrm{el}}$ values for the $^{17}$Ne + $^{208}$Pb system, we employ the OM method with the Schr\"odinger equation as follows~\cite{maha86,kim02,so04}:
\begin{equation} \label{scodinger}
[E - T_{l} (r)] \chi^{(+)}_{l} (r) = U_{\textrm{OM}} (r)~\chi^{(+)}_{l} (r)
\end{equation}
where $T_{l} (r)$ and $\chi^{(+)}_{l} (r)$ represent the kinetic energy operator and the distorted partial wave function expressed as a function of the angular momentum $l$, respectively. The OM potential $U_{\textrm{OM}} (r)$ in Eq.~(\ref{scodinger}) comprises the real monopole Coulomb potential
$V_{\textrm{C}} (r)$, the energy-independent real bare potential $V_{0} (r)$, and the energy-dependent complex dynamic polarization potential (DPP) $U_{\textrm{DPP}} (r)$ as follows~\cite{heo22}:
\begin{eqnarray} \label{potential}
U_{\textrm{OM}} (r) &=& V_{\textrm{C}} (r) - V_{0} (r) - U_{\textrm{DPP}} (r) \nonumber \\
                    &=& V_{\textrm{C}} (r) - V_{0} (r) - U_{\textrm{DR}} (r) - U_{\textrm{F}} (r) \nonumber \\
                    &=& V_{\textrm{C}} (r) - V_{0} (r) - U_{\textrm{inel}} (r) - U_{\textrm{br}} (r) - U_{\textrm{F}} (r) \nonumber \\
                    &=& V_{\textrm{C}} (r) - V_{0} (r) - i W^{\textrm{C}}_{\textrm{inel}} (r) - U^{\textrm{N}}_{\textrm{inel}} (r) - i W^{\textrm{C}}_{\textrm{br}} (r) - U^{\textrm{N}}_{\textrm{br}} (r) - U_{\textrm{F}} (r).
\end{eqnarray}

\begin{figure}
\begin{tabular}{c}
\includegraphics[width=1.00\linewidth]{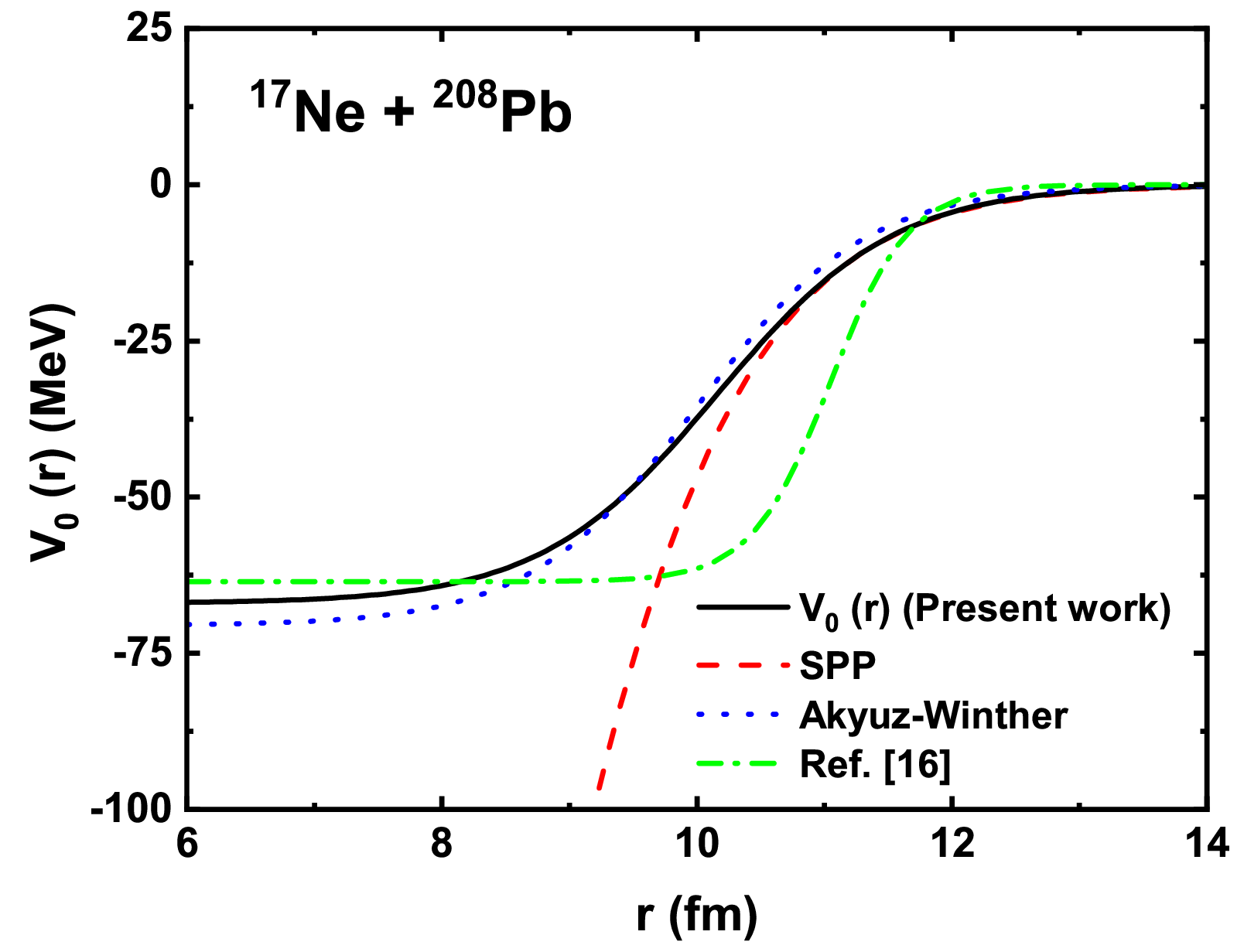}
\end{tabular}
\caption{Bare potentials for the $^{17}$Ne + $^{208}$Pb system. The dashed red line shows the SPP made using nuclear densities for both $^{17}$Ne and $^{208}$Pb nuclei. The solid black line represent a newly parameterized SPP potential in the Woods-Saxon form, obtained through the $\chi^{2}$-fitting. The dotted blue and dash-dotted green lines represent the potentials, respectively, as provided by Akyuz-Winther~\cite{akyu79} and Ref.~\cite{ovej23}.}
\label{bare-pot}
\end{figure}

In Eq.~(\ref{potential}), $ V_{0} (r)$ denotes the real bare potential with a volume-type Woods-Saxon potential, which is an independent potential for the incident energy of the projectile and is expressed as
\begin{equation} \label{bare}
V_{0} (r) = V_{0} \Big[1+\exp (\frac{r-R_{\textrm{0}}}{a_{\textrm{0}}})\Big]^{-1}
\end{equation}
where $R_{\textrm{0}}$ = $r_{\textrm{0}}$ (${A_{1}^{1/3}}$ + ${A_{2}^{1/3}}$). $A_{1}$ and $A_{2}$ represent the mass numbers of the projectile and target nuclei, respectively. For reference, this real bare potential is created by reconstituting the S\~{a}o Paulo potential (SPP) as follows~\cite{cham02}. As done in Refs.~\cite{choi18,choi21}, we first construct the SPP (the dashed red line) using the nuclear densities for both $^{17}$Ne and $^{208}$Pb nuclei, as shown in Fig.~\ref{bare-pot}. However, the SPP based on the folding potential has a considerably large depth parameter, as shown in Fig.~\ref{bare-pot}. To solve this problem, we newly parameterize the SPP potential into the Woods-Saxon form through a $\chi^{2}$-fitting. This is performed to effectively match at $r$ $\geq$ 11 fm, where a strong absorption reaction occurs.
Consequently, we obtain a new Woods-Saxon type potential (bare potential), denoted by the solid black line in Fig. ~\ref{bare-pot}. In the near-barrier incident energy region, when the projectile and target nucleus come close together, the nuclear reaction at the surface becomes crucial because the two nuclei cannot collide deeply. Therefore, we consider the $\chi^{2}$-fitting more precisely to the outside than the inside of the potential. To compare the newly obtained bare potential with that provided by Akyuz-Winther (AW)~\cite{akyu79} and Ref.~\cite{ovej23} (dotted blue and dash-dotted green lines), these potentials are drawn together in Fig.~\ref{bare-pot}. The potential parameters are listed in Table~\ref{bare_para_17ne}. Interestingly, our bare potential and the AW potential are similar, as shown in Fig.~\ref{bare-pot} and Table~\ref{bare_para_17ne}.

\begin{table}[H]
\begin{center}
\caption{Parameter sets of the bare potentials for the $^{17}$Ne + $^{208}$Pb system.}
\label{bare_para_17ne}
\begin{ruledtabular}
\begin{tabular}{cccc}
Model             & $V_{0}$                                & $a_{0}$                               & $r_{0}$                   \\
                  & (MeV)                                  & (fm)                                  & (fm)                      \\ \hline
Bare (present)    & 67.0                                   & 0.692                                 & 1.196                     \\
AW~\cite{akyu79}  & 70.6                                   & 0.660                                 & 1.178                     \\
Ref.~\cite{ovej23}& 63.5                                   & 0.309                                 & 1.300                     \\
\end{tabular}
\end{ruledtabular}
\end{center}
\end{table}

According to the incident energy of the projectile, the DPP denoted as $U_{\textrm{DPP}} (r)$ can be segmented into the DR and fusion reaction potentials. In this research, we divide the DR channels into the inelastic scattering and breakup reaction channels. More specifically, we express the potentials related to these channels as $U_{\textrm{inel}} (r)$ and $U_{\textrm{br}} (r)$, respectively, in Eq.~(\ref{potential}).

Considering Refs.~\cite{ovej23,chro02,marg16}, $^{17}$Ne has no excited state below the two-proton separation energy ($S_{2p}$ = 0.944 MeV). Instead, there are three low-lying resonance states, which are $\frac{3}{2}^{-}$ state [$E_{x}$ = 1.288 MeV and $B(E2)$ = $95^{+26}_{-36} \textrm{e}^{2}\textrm{fm}^{4}$ or $66^{+18}_{-25} \textrm{e}^{2}\textrm{fm}^{4}$], $\frac{5}{2}^{-}$ state [$E_{x}$ = 1.764 MeV and $B(E2)$ = $90 \pm 18 \textrm{e}^{2}\textrm{fm}^{4}$ or $179 \pm 26 \textrm{e}^{2}\textrm{fm}^{4}$], and $\frac{3}{2}^{-}$ state [$E_{x}$ = 2.692 MeV and $B(E2)$ = $69 \pm 10 \textrm{e}^{2}\textrm{fm}^{4}$], above two proton separation energy. Here, $E_{x}$ is the excitation energy. In the $\frac{3}{2}^{-}$ [$E_{x}$ = 1.288 MeV] state, two decay modes are possible by either concurrent 2$p$ emission or gamma-ray emission. However, because the probability of simultaneous 2$p$ emissions is less than 0.77\%, it has been reported that only gamma emissions are observed. Thus, the channel owing to the coupling between this excited state ($\frac{3}{2}^{-}$) and the ground state ($\frac{1}{2}^{-}$) is regarded as an inelastic scattering channel in this study. However, for the $\frac{5}{2}^{-}$ state, Ref.~\cite{chro02} reported that no evidence of gamma emission or simultaneous 2$p$ emission was measured experimentally. Instead, $^{17}$Ne projectile decayed to $^{15}$O nucleus through two sequential proton emissions. Moreover, the $\frac{3}{2}^{-}$ state with $E_{x}$ = 2.692 MeV also had the two sequential proton emission modes, as in the case of $\frac{5}{2}^{-}$ state. Thus, these two channels ($\frac{5}{2}^{-}$ state with $E_{x}$ = 1.764 MeV and $\frac{3}{2}^{-}$ state with $E_{x}$ = 2.692 MeV) are associated with the inclusive $^{15}$O angular distribution and are regarded as breakup channels and not as inelastic scattering channels in this study.

To consider the Coulomb excitation effect for the inelastic scattering and/or breakup channel, we introduced the Coulomb excitation potential. This involves various interactions such as Coulomb dipole excitation (CDE; E1 transition) and Coulomb quadrupole excitation (CQE; E2 transition). However, as mentioned above, the interaction in the case of the inelastic scattering and/or breakup channel of the $^{17}$Ne projectile is related to the E2 transition. Thus, we only need to consider the CQE interaction with an imaginary potential, as described below~\cite{love77}:
\begin{equation} \label{cqe}
                 W^{\textrm{C}}_{\textrm{inel.}~(\textrm{br})} (r) = \left \{
                           \begin{array}{ll}
-\Big[1-\frac{2}{7}\Big(\frac{{r_{C}}}{r}\Big)^{2}-\frac{1}{21}\Big(\frac{r_{C}}{r}\Big)^{4}\Big]
\Big[1-\Big(\frac{Z_{1} Z_{2} e^{2}}{rE_{c.m.}}\Big) \Big]^{-1/2}~
\frac{W_{P}}{r^{5}} &~ \mbox{for}~r \geq r_{C}   \vspace{2ex} \\
-\frac{2\sqrt{10}}{3}~\frac{W_{P} r^{4}}{r^{9}_{C}}  &~ \mbox{for}~ r <  {r_{C}}    \\
                           \end{array}
                 \right.
\end{equation}
with
\begin{equation} \label{wp}
W_{p} = 0.01676 \frac {\mu Z_{2}^{2}}{k} B(E2) g_{2} (\xi).
\end{equation}
Here, $B(E2)$ represents the quadrupole electric transition strength related to the E2 transition, and $g_{2} (\xi)$ indicate an adiabaticity correction factor given as a function of the adiabaticity parameter $\xi$ = $a_{0} \varepsilon$/$\hbar v$~\cite{heo22,andr94,love77}  with the distance of the closest approach $a_{0}$. $Z_{2}$ denotes the charge number of the target nucleus.

To investigate the breakup effect of the $^{17}$Ne + $^{208}$Pb system, the remaining DR channels must be considered, excluding the inelastic scattering channels. Similar to the inelastic scattering mentioned earlier, the breakup effect can be classified into nuclear interaction and Coulomb excitation components because it mainly occurs on the surface or at a long region distance caused by the Coulomb and nuclear forces acting between the projectile and target nucleus. To consider the breakup effect of the Coulomb interaction mentioned above, the CQE potential in Eq.~(\ref{cqe}) can be applied as is. To explain the breakup effect occurring near the nuclear surface via nuclear interaction, we manipulated the nuclear interaction component of the $U_{\textrm{br}}(r)$ potential as another Wood-Saxon potential with the following surface-type constituent~\cite{rehm82}:
\begin{equation} \label{surface-woods}
U^{\textrm{N}}_{\textrm{br}}(r)= -4~a_{\textrm{i,~br}}^{\textrm{N}}~(V^{\textrm{N}}_{\textrm{br}} + i W^{\textrm{N}}_{\textrm{br}})
~\frac{d[1+\exp(X^{\textrm{N}}_{\textrm{i,~br}})]^{-1}}{dr},~~~~\textrm{i} = V~\textrm{and}~W,
\end{equation}
where $X^{\textrm{N}}_{\textrm{i,~br}}$ = $(r-R^{\textrm{N}}_{\textrm{i,~br}})/a^{\textrm{N}}_{\textrm{i,~br}}$
with $R^{\textrm{N}}_{\textrm{i,~br}}$ = $r^{\textrm{N}}_{\textrm{i,~br}}$ (${A_{1}^{1/3}}$ + ${A_{2}^{1/3}}$).
Note that the potential form used to explain the breakup reaction by the nuclear interaction is similar to the differential form of the Woods-Saxon type used in describing inelastic scattering by the nuclear interaction.

Ultimately, the fusion reaction of the DPP primarily occurred not on the nucleus surface but rather deep inside.
Therefore, the fusion potential is considered herein as a volume-type Woods--Saxon potential:
\begin{equation} \label{volume-woods}
U_{\textrm{F}} (r) = (V_{\textrm{F}} + i W_{\textrm{F}})~[1+\exp(X_{\textrm{i,~F}})]^{-1},~~~~\textrm{i} = V~\textrm{and}~W,
\end{equation}
where $X_{\textrm{i,~F}}$ = $(r-R_{\textrm{i,~F}})/a_{\textrm{i,~F}}$ with $R_{\textrm{i,~F}}$ = $r_{\textrm{i,~F}}$ (${A_{1}^{1/3}}$ + ${A_{2}^{1/3}}$). In this study, the meaning of fusion ``F" is formed by the total fusion combining the complete and incomplete fusion.

\section{Results}
\label{simple approach}
\subsection{OM calculations}
\begin{figure}
\begin{tabular}{c}
\includegraphics[width=0.495\linewidth]{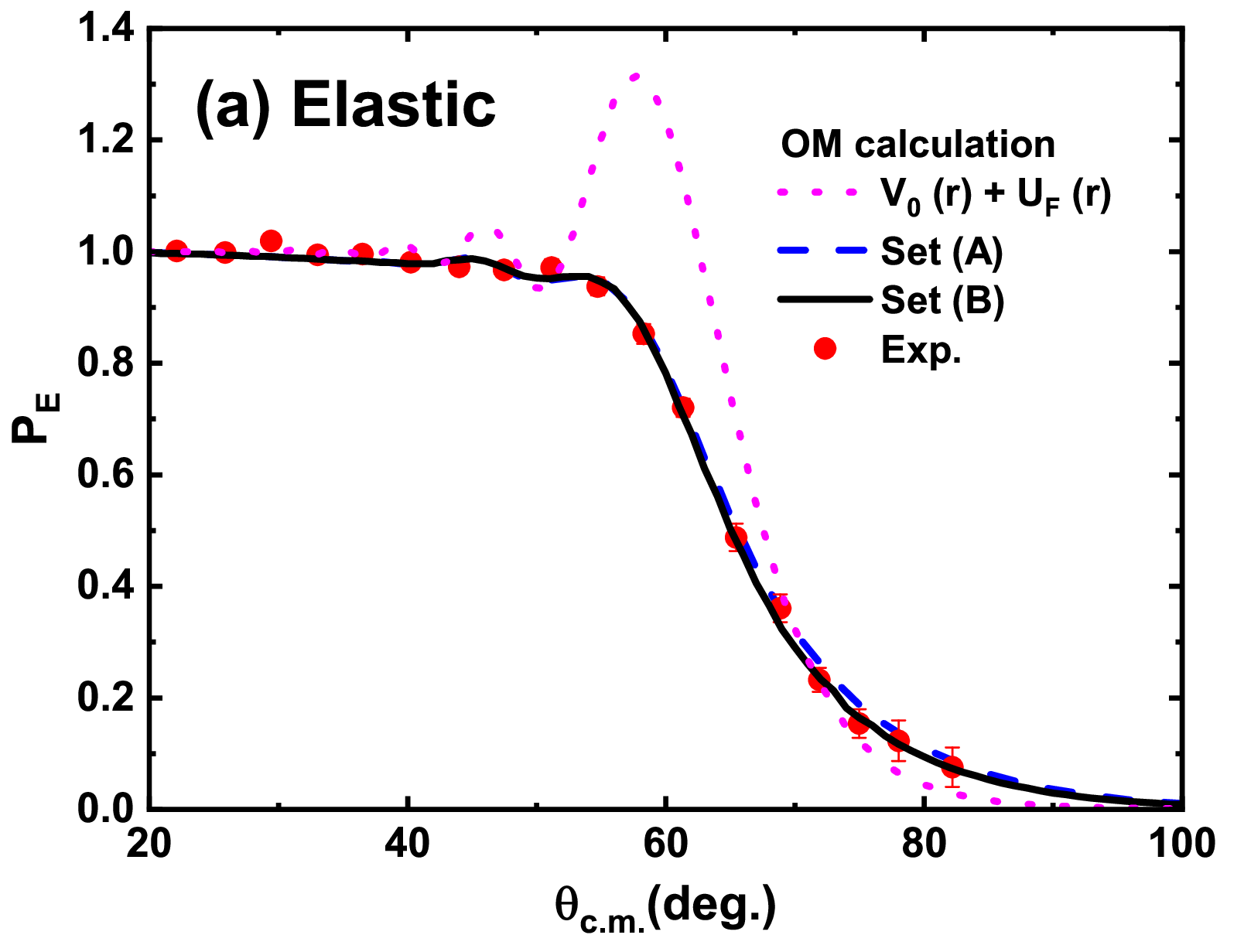}
\includegraphics[width=0.495\linewidth]{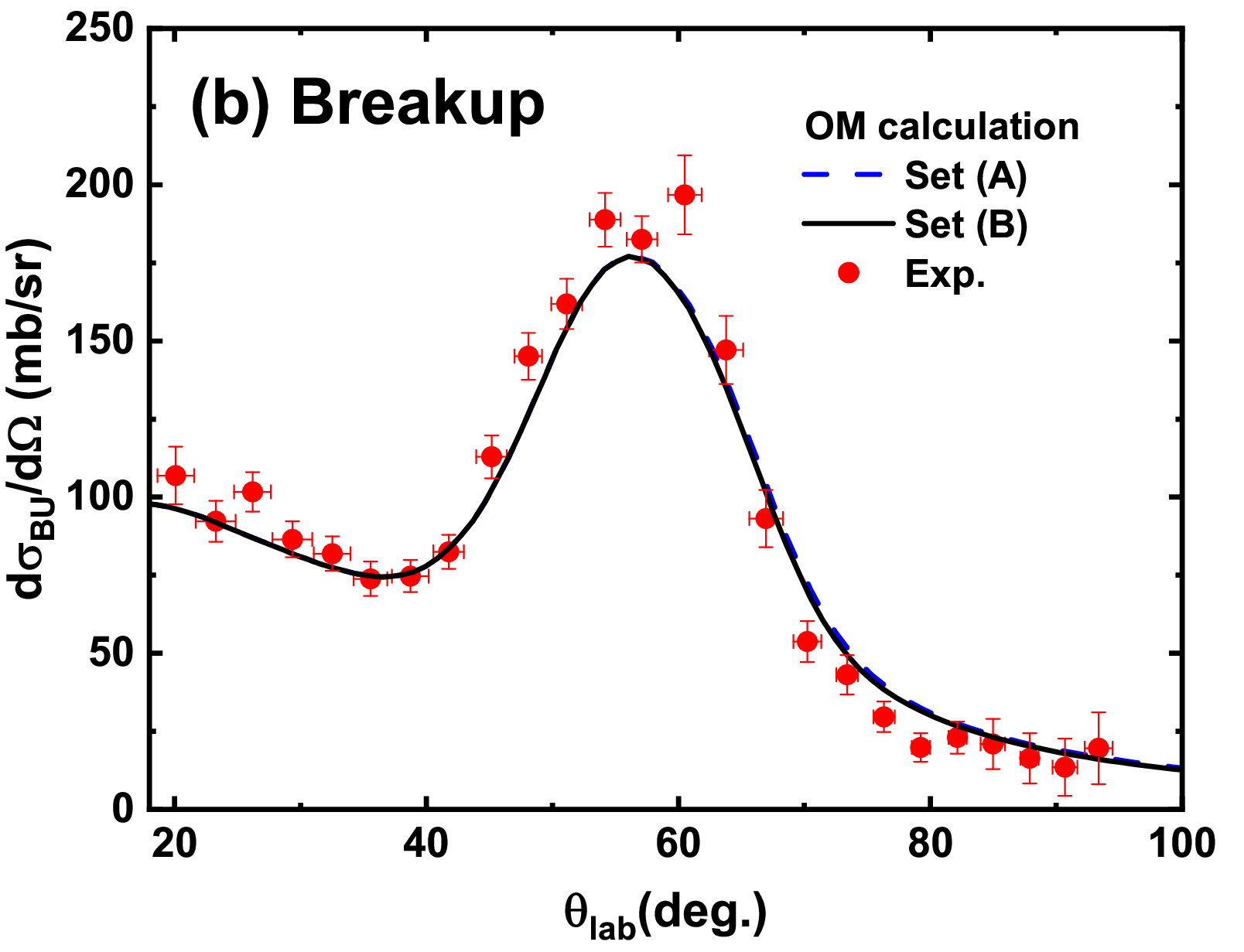}
\end{tabular}
\caption{(a) $P_{\textrm{el}}$ ratios and (b) angular distributions of the breakup ($^{15}$O production) cross section for the $^{17}$Ne + $^{208}$Pb system. The dashed blue and solid black lines denote the theoretical ratios $P_{\textrm{el}}$ and angular distributions using the parameter sets (A) and (B) in Table~\ref{parameters_17ne}, respectively. The solid red circles denote the experimental data at $E_{\textrm{lab}}$ = 136 MeV taken from Ref.~\cite{ovej23}. Note that the dotted magenta line indicates $P_{\textrm{el}}$ ratio only based on the $V_{0}$ + $U_{\textrm{F}}$ in Eq.~(\ref{potential}), where $U_{\textrm{F}}$ has been extracted through the set(B). To perform these calculations, we used the $\frac{3}{2}^{-}$ state [$E_{x}$ = 1.288 MeV and $B(E2)$ = 41 $\textrm{e}^{2}\textrm{fm}^{4}$] for the inelastic interaction and the $\frac{5}{2}^{-}$ state [$E_{x}$ = 1.764 MeV and $B(E2)$ = 72 $\textrm{e}^{2}\textrm{fm}^{4}$] and $\frac{3}{2}^{-}$ state [$E_{x}$ = 2.692 MeV and $B(E2)$ = 59 $\textrm{e}^{2}\textrm{fm}^{4}$] for the breakup one. Note that these values correspond to the minimum value of $B(E2)$ introduced in Refs.~\cite{ovej23,chro02,marg16}.}
\label{om_cal}
\end{figure}

As depicted in the dotted magenta line in Fig.~\ref{om_cal} (a), the experimental $P_{\textrm{el}}$ ratio of the $^{17}$Ne + $^{208}$Pb system cannot be reproduced through a potential defined as the sum of $V_{0} (r)$ and $U_{\textrm{F}} (r)$ in Eq. ~(\ref{potential}). This indicates that the additional absorption channels are open near the Coulomb barrier energy region. Therefore, the additional DPPs are used to reproduce the experimental results. This study considers inelastic scattering, breakup reactions, and fusion reactions as the additional DPPs.

In principle, to explain the elastic scattering data for the $^{17}$Ne + $^{208}$Pb system using these DPPs, various experimental data are required for the input data. However, in the $^{17}$Ne + $^{208}$Pb system, there are two sets of experimental data: the elastic scattering and breakup cross-section data. Nevertheless, we can calculate the total fusion cross sections between the incident isotopes and target nucleus through the Barrier penetration model (BPM)~\cite{hagi99}. The extracted total fusion cross section obtained from the BPM using $V_{0} (r)$ in Eq.~(\ref{potential}) is approximately 994.3 mb.

Utilizing the experimental elastic scattering, experimental breakup reaction, and theoretical total fusion cross-section data, we performed the $\chi^{2}$ analysis.
As discussed earlier, breakup reactions occur primarily near the surface of the projectile and the target nucleus; therefore, we employed the differential form of the Woods-Saxon potential in the calculations presented in Eq.~(\ref{surface-woods}).
Consequently, the $\chi^{2}$ fitting could be performed on twelve adjustable parameters ($V^{\textrm{N}}_{\textrm{br}}$, $W^{\textrm{N}}_{\textrm{br}}$, $a^{\textrm{N}}_{\textrm{v, br}}$, $a^{\textrm{N}}_{\textrm{w, br}}$, $r^{\textrm{N}}_{\textrm{v, br}}$, $r^{\textrm{N}}_{\textrm{w, br}}$, $V_{\textrm{F}}$, $W_{\textrm{F}}$, $a_{\textrm{v, F}}$, $a_{\textrm{w, F}}$, $r_{\textrm{v, F}}$, $r_{\textrm{w, F}}$) in $U^{\textrm{N}}_{\textrm{br}} (r)$ and $U_{\textrm{F}} (r)$. However, as specific adjustable parameters were entangled, we reduced the parameters in the $\chi^{2}$ fitting by applying it on eight adjustable parameters ($V^{\textrm{N}}_{\textrm{br}}$, $W^{\textrm{N}}_{\textrm{br}}$, $a^{\textrm{N}}_{\textrm{br}}$, $r^{\textrm{N}}_{\textrm{br}}$, $V_{\textrm{F}}$, $W_{\textrm{F}}$, $a_{\textrm{F}}$, $r_{\textrm{F}}$) by assuming the diffuseness and radius parameters as $a^{\textrm{N}}_{\textrm{V, br}}$ = $a^{\textrm{N}}_{\textrm{w, br}}$ $\equiv$ $a^{\textrm{N}}_{\textrm{br}}$, $a_{\textrm{v, F}}$ = $a_{\textrm{w, F}}$ $\equiv$ $a_{\textrm{F}}$, $r^{\textrm{N}}_{\textrm{v, br}}$ = $r^{\textrm{N}}_{\textrm{w, br}}$ $\equiv$ $r^{\textrm{N}}_{\textrm{br}}$, and $r_{\textrm{v, F}}$ = $r_{\textrm{w, F}}$ $\equiv$ $r_{\textrm{F}}$.

The parameters obtained from $\chi^{2}$ fitting are listed in set (A) of Table~\ref{parameters_17ne}.
Based on these parameters, we plotted the theoretical $P_{\textrm{el}}$ ratio (i.e., the ratio of the elastic scattering cross section to the Rutherford cross section) and the angular distribution of the breakup cross section as the dashed blue line in Fig.~\ref{om_cal}. Overall, the theoretical $P_{\textrm{el}}$ ratio and the angular distribution of the breakup cross section are consistent with the experimental ratio, except for one or two points. Interestingly, the rainbow peak observed around $\theta_{\textrm{c.m.}}$ = 60$^{\circ}$ in the dotted magenta line is strongly suppressed owing to the effect of inelastic scattering and breakup reactions. This can be attributed to the $^{17}$Ne nucleus, which exhibits relatively low values for its first excited energy ($E^{1st}_{x}$ = 1.288MeV) and separation energies ($S_{2p}$=0.944 and $S_{p}$ = 1.464 MeV). These similarities are reminiscent of weakly bound $^{11}$Li and $^{11}$Be nuclei.

To investigate the angular distribution of the breakup cross section, we utilize the following form ~\cite{kim02} for reference~\footnote{Our previous papers~\cite{heo22,so16} have a typo. The equation (16) in Ref~\cite{heo22} and Eq.~(5) in Ref.~\cite{so16} is to be replaced by Eq.~(\ref{differ_cross}) in the present paper.}:
\begin{eqnarray}
\textcolor[rgb]{0.00,0.00,0.00}{\frac{d\sigma_{\textrm{br}}}{d\Omega}}&\textcolor[rgb]{0.00,0.00,0.00}{=}&\textcolor[rgb]{0.00,0.00,0.00}{\frac{ka_{0}}{16\pi}\frac{1}{\cos(\frac{\theta_{\textrm{c.m.}}}{2})\sin^{3}(\frac{\theta_{\textrm{c.m.}}}{2})}~\sigma_{\textrm{br}; l}} \nonumber \\
&\textcolor[rgb]{0.00,0.00,0.00}{=}&\textcolor[rgb]{0.00,0.00,0.00}{\frac{ka_{0}}{16\pi}\frac{1}{\cos(\frac{\theta_{\textrm{c.m.}}}{2})\sin^{3}(\frac{\theta_{\textrm{c.m.}}}{2})}~\frac{\pi}{k^{2}}(2l + 1)~T_{\textrm{br}; l},}
\label{differ_cross}
\end{eqnarray}
with
\begin{equation}
T_{\textrm{br}; l} = \frac{8}{\hbar v} \int^{\infty}_{0} |\chi^{+}_{l} (r)|^{2}~[W_{\textrm{br}} (r)]dr = \frac{8}{\hbar v} \int^{\infty}_{0} |\chi^{+}_{l} (r)|^{2}~[W^{\textrm{N}}_{\textrm{br}} (r) + W^{\textrm{C}}_{\textrm{br}} (r)]dr.
\label{T-inel}
\end{equation}
\textcolor[rgb]{0.00,0.00,0.00}{The point to note is that the orbital angular momentum ($l$) is treated as an integer value in quantum-mechanical expression. However, according to the equation for $l$ and $\theta_{\textrm{c.m.}}$~\cite{kim02}, $l$ $\equiv$ $l_{\theta_{\textrm{c.m.}}}$ = $\frac{ka_{0}}{2}$ $\cot(\frac{\theta_{\textrm{c.m.}}}{2})$ used in the present research, the orbital angular momentum ($l _{\theta_{\textrm{c.m.}}}$) is obtained by continuous scattering angles ($\theta_{\textrm{c.m.}}$), and does not have discrete integer values, but rather has continuous values. To address this issue, we have exploited semi-classical approximation methods customarily used in several semi-classical treatments of the quantum mechanical Rutherford cross section~\cite{bass80,satc80,mott65,ford59}. That is, since the main contribution to the differential cross sections stems from the partial waves in the vicinity of $l$ $\approx$ $l_{\theta_{\textrm{c.m.}}}$, for a given scattering angle $\theta_{\textrm{c.m.}}$, one can found the nearest $l$ with the integer values and calculate Eq.~(\ref{T-inel}) using the $l$. In order to obtain more reasonable calculations in the present work, however, we have calculated $\sigma_{\textrm{br}; l_{\theta_{\textrm{c.m.}}}}$ by taking an interpolation of the results obtained by the two partial wave cross sections ($\sigma_{\textrm{br}; l}$ and $\sigma_{\textrm{br}; l+1}$) corresponding to the two nearby $l$ and $l+1$ for $l$ $<$ $l_{\theta_{\textrm{c.m.}}}$ $<$ $l+1$~($l$ is the integer value). Of course, this procedure is most justified when the scattering follows the classical path. Detailed discussion was also done in the previous paper done by one of present authors~\cite{kim02}.}

Furthermore, we evaluated the inelastic, breakup, and fusion cross sections ($\sigma_{\textrm{inel.}}$, $\sigma_{\textrm{br}}$, and $\sigma_{\textrm{F}}$) can be expressed as~\cite{udag84,huss84}
\begin{equation}
\sigma_{\textrm{i}} = \frac{2}{\hbar v} <\chi^{+} (r)|~W_{\textrm{i}} (r)|~\chi^{+} (r)>~~~~\textrm{i} = \textrm{inel.,~br,~and~F}.
\label{T-br}
\end{equation}
These parameters are listed in set (A) of Table~\ref{cross-sec}. Notably, we used $W_{\textrm{i}} (r)$ in Eq. ~(\ref{T-br}) as $W_{\textrm{inel.}} (r)$ = $W^{\textrm{N}}_{\textrm{inel.}} (r)$ + $W^{\textrm{C}}_{\textrm{inel.}} (r)$, $W_{\textrm{br}} (r)$ = $W^{\textrm{N}}_{\textrm{br}} (r)$ + $W^{\textrm{C}}_{\textrm{br}} (r)$, and $W_{\textrm{F}} (r)$ for $\sigma_{\textrm{inel.}}$, $\sigma_{\textrm{br}}$, and $\sigma_{\textrm{F}}$, respectively.

\begin{table}
\begin{center}
\caption{Optimized parameter sets of DPPs for the $^{17}$Ne + $^{208}$Pb system.}
\label{parameters_17ne}
\begin{ruledtabular}
\begin{tabular}{cccccccccccccc}
Set&\multicolumn{4}{c}{Inelastic scattering}& \multicolumn{4}{c}{Breakup}           & \multicolumn{4}{c}{Fusion}      & $\chi^{2}$\\  \hline
   & $V^{\textrm{N}}_{\textrm{inel}}$       & $W^{\textrm{N}}_{\textrm{inel}}$      & $a^{\textrm{N}}_{\textrm{inel}}$
   & $r^{\textrm{N}}_{\textrm{inel}}$       & $V^{\textrm{N}}_{\textrm{br}}$        & $W^{\textrm{N}}_{\textrm{br}}$
   & $a^{\textrm{N}}_{\textrm{br}}$         & $r^{\textrm{N}}_{\textrm{br}}$        & $V_{\textrm{F}}$
   & $W_{\textrm{F}}$                       & $a_{\textrm{F}}$                      & $r_{\textrm{F}}$                &           \\
   & (MeV)                                  & (MeV)                                 & (fm)
   & (fm)                                   & (MeV)                                 & (MeV)
   & (fm)                                   & (fm)                                  & (MeV)
   & (MeV)                                  & (fm)                                  & (fm)                  &  \\ \cline{2-5}\cline{6-9}\cline{10-13}
(A)&  --                                    &  --                                   &  --
   &  --                                    & 0.354                                 & 0.142
   & 0.760                                  & 1.627                                 & -5.164
   & 7.305                                  & 0.222                                 & 1.44                       & 2.6  \\
(B)&  -2.826                                & 1.305                                 & 0.178
   &  1.451                                 & 0.354                                 & 0.142
   & 0.760                                  & 1.627                                 & 46.088
   & 94.720                                 & 0.340                                 & 1.28                       & 2.1  \\
\end{tabular}
\end{ruledtabular}
\end{center}
\end{table}

\begin{table}
\caption{Inelastic scattering, breakup, total fusion, and total reaction cross sections for the $^{17}$Ne + $^{208}$Pb system at $E_{\textrm{lab}}$ = 136 MeV. Sets (A) and (B) are the results obtained using the sets listed in Table~\ref{parameters_17ne}.}
\label{cross-sec}
\begin{ruledtabular}
\begin{tabular}{cccccccc}
                         &       set    & \multicolumn{2}{c}{$\sigma_{\textrm{inel.}}$}         & \multicolumn{2}{c}{$\sigma_{\textrm{br}}$}  & $\sigma_{\textrm{F}}$  & $\sigma_{\textrm{R}}$ \\ \cline{3-4} \cline{5-6}
                         &              & $\sigma^{\textrm{N}}_{\textrm{inel}}$  &  $\sigma^{\textrm{C}}_{\textrm{inel}}$
                                        & $\sigma^{\textrm{N}}_{\textrm{br}}$    & $\sigma^{\textrm{C}}_{\textrm{br}}$
                                        &                                        &                       \\
                         &              &      (mb)    &    (mb)    &    (mb)    &    (mb)    &     (mb)    &    (mb)     \\ \hline
Present work (OM)        &      (A)     &       --     &   130.4    &    320.0   &    282.7   &   1126.8    &         1859.9        \\
Present work (OM)        &      (B)     &     250.9    &   128.3    &    316.4   &    278.1   &    915.6    &         1889.3        \\
Ref.~\cite{ovej23} (OM)  &              &       --     &     --     &     --     &     --     &     --      &         1800.0        \\
Ref.~\cite{ovej23} (CC)  &              & \multicolumn{2}{c}{350.0} &     --     &     --     &     --      &         1903.0        \\
\end{tabular}
\end{ruledtabular}
\end{table}

Based on the calculations performed thus far, we determined the contribution of the elastic scattering, inelastic scattering, breakup, and total fusion cross sections of the $^{17}$Ne + $^{208}$Pb system, as shown in set (A) of Table~\ref{cross-sec}. However, owing to the absence of the experimental inelastic scattering cross-section data, the contribution of each component (nuclear and Coulomb) to the inelastic scattering cross section could not be adequately investigated. As inelastic scattering channel can also be caused by the nuclear and Coulomb interactions, we must also consider the inelastic scattering channel by the nuclear interaction. In particular, the absence of the nuclear component of the inelastic scattering interaction suggests that the flux corresponding to this interaction has flowed into other reaction channels. In fact, as shown in set (A) of Table~\ref{cross-sec}, the total fusion cross section obtained in our calculation (1126.8 mb) is approximately 132.5 mb larger than that obtained in the CC calculation (994.3 mb). To isolate the nuclear component of the inelastic scattering interaction, we employ a new interaction. As inelastic scattering primarily occurred near the nucleus surface, the nuclear component of the inelastic scattering potential was generally considered using a differential form of the Woods-Saxon potential, as expressed in Ref.~\cite{rehm82}:
\begin{equation} \label{inel_pot}
U^{\textrm{N}}_{\textrm{inel.}}(r)= -4~a_{\textrm{i,~inel}}^{\textrm{N}}~~(V^{\textrm{N}}_{\textrm{inel}} + i W^{\textrm{N}}_{\textrm{inel}})
~\frac{d[1+\exp(X^{\textrm{N}}_{\textrm{i,~inel}})]^{-1}}{dr},~~~~\textrm{i} = V~\textrm{and}~W,
\end{equation}
where $X^{\textrm{N}}_{\textrm{i,~inel}}$ = $(r-R^{\textrm{N}}_{\textrm{i,~inel}})/a^{\textrm{N}}_{\textrm{i,~inel}}$
where $R^{\textrm{N}}_{\textrm{i,~inel}}$ = $r^{\textrm{N}}_{\textrm{i,~inel}}$ (${A_{1}^{1/3}}$ + ${A_{2}^{1/3}}$).

With the differential form of the Woods-Saxon potential corresponding to the nuclear component of the inelastic scattering potential, we performed the additional $\chi^{2}$ fitting using $U_{\textrm{DPP}} (r)$ = $U^{\textrm{N}}_{\textrm{inel}} (r)$ + $i W^{\textrm{C}}_{\textrm{inel}} (r)$ + $U^{\textrm{N}}_{\textrm{br}} (r)$ + $U^{\textrm{C}}_{\textrm{br}} (r)$ + $U_{\textrm{F}} (r)$ in Eq. ~(\ref{potential}). For reference, $W^{\textrm{C}}_{\textrm{inel}} (r)$ and $U^{\textrm{C}}_{\textrm{br}} (r)$ potentials related to the CQE transition did not require adjustable parameters. In principle, we should perform $\chi^{2}$ fitting on eighteen adjustable parameters ($V^{\textrm{N}}_{\textrm{inel}}$, $W^{\textrm{N}}_{\textrm{inel}}$, $a^{\textrm{N}}_{\textrm{v, inel}}$, $a^{\textrm{N}}_{\textrm{w, inel}}$, $r^{\textrm{N}}_{\textrm{v, inel}}$, $r^{\textrm{N}}_{\textrm{w, inel}}$, $V^{\textrm{N}}_{\textrm{br}}$, $W^{\textrm{N}}_{\textrm{br}}$, $a^{\textrm{N}}_{\textrm{v, br}}$, $a^{\textrm{N}}_{\textrm{w, br}}$, $r^{\textrm{N}}_{\textrm{V, br}}$, $r^{\textrm{N}}_{\textrm{w, br}}$, $V_{\textrm{F}}$, $W_{\textrm{F}}$, $a_{\textrm{v, F}}$, $a_{\textrm{w, F}}$, $r_{\textrm{v, F}}$, $r_{\textrm{w, F}}$) in $U^{\textrm{N}}_{\textrm{inel}} (r)$,  $U^{\textrm{N}}_{\textrm{br}} (r)$, and $U_{\textrm{F}} (r)$. Because we have already determined the parameters related to the breakup reaction potential that effectively reproduce the experimental breakup cross-section data, we can fix them ($V^{\textrm{N}}_{\textrm{br}}$, $W^{\textrm{N}}_{\textrm{br}}$, $a^{\textrm{N}}_{\textrm{br}}$, $r^{\textrm{N}}_{\textrm{br}}$) obtained through the set (A) of Table~\ref{parameters_17ne}. To reduce these parameters, we assume that the diffuseness and radius parameters of the inelastic scattering and total fusion potential are $a^{\textrm{N}}_{\textrm{v, inel.}}$ = $a^{\textrm{N}}_{\textrm{w, inel.}}$ $\equiv$ $a^{\textrm{N}}_{\textrm{inel.}}$,
$r^{\textrm{N}}_{\textrm{v, inel.}}$ = $r^{\textrm{N}}_{\textrm{w, inel.}}$ $\equiv$ $r^{\textrm{N}}_{\textrm{inel.}}$, $a_{\textrm{v, F}}$ = $a_{\textrm{w, F}}$ $\equiv$ $a_{\textrm{F}}$ and $r_{\textrm{v, F}}$ = $r_{\textrm{w, F}}$ $\equiv$ $r_{\textrm{F}}$. the radius parameters. Thereafter, we perform a second $\chi^{2}$ fitting with eight adjustable parameters ($V^{\textrm{N}}_{\textrm{inel.}}$, $W^{\textrm{N}}_{\textrm{inel.}}$, $a^{\textrm{N}}_{\textrm{inel.}}$, $r^{\textrm{N}}_{\textrm{inel.}}$, $V^{\textrm{N}}_{\textrm{F}}$, $W^{\textrm{N}}_{\textrm{F}}$, $a^{\textrm{N}}_{\textrm{F}}$, $r^{\textrm{N}}_{\textrm{F}}$).
The parameters obtained from $\chi^{2}$ fitting are listed in set (B) of Table~\ref{parameters_17ne}.

Based on parameter set (B) of Table~\ref{parameters_17ne}, we plotted the theoretical $P_{\textrm{el}}$ ratio and the angular distribution of the breakup cross section emerging from the nuclear component for inelastic scattering interaction, as shown by the solid black line in Fig. ~{\ref{om_cal}}. In these calculations, the theoretical $P_{\textrm{el}}$ ratio and angular distribution of the breakup cross section were consistent with the experimental data for almost all scattering angles, except for one or two points. As shown in set (B) of Table~\ref{cross-sec}, part of the fusion cross section shifted to inelastic scattering because of the addition of a new nuclear component of the inelastic scattering interaction, as expected.

Finally, we calculated the inelastic scattering, breakup reaction, and fusion cross sections expressed in Eq. (\ref{T-br}), which are listed in set (B) of Table~\ref{cross-sec}. The results showed that the inelastic scattering cross sections exhibited a greater contribution through the nuclear interaction than through Coulomb excitation. By contrast, the breakup reaction cross section showed that the contributions of the two interactions were similar. Interestingly, the contribution of the DR cross section (i.e., the sum of the inelastic scattering and breakup reaction cross sections) to the total reaction was approximately half ($\sigma_{\textrm{DR}}$/$\sigma_{\textrm{R}}$ $\approx$ 0.52) of the total reaction cross section, as shown in set (B) of Table~\ref{cross-sec}. This implies that a significant portion of the elastic scattering channel is converted to the DR channel through processes, such as inelastic scattering and breakup reaction.

\begin{figure}
\begin{tabular}{c}
\includegraphics[width=0.495\linewidth]{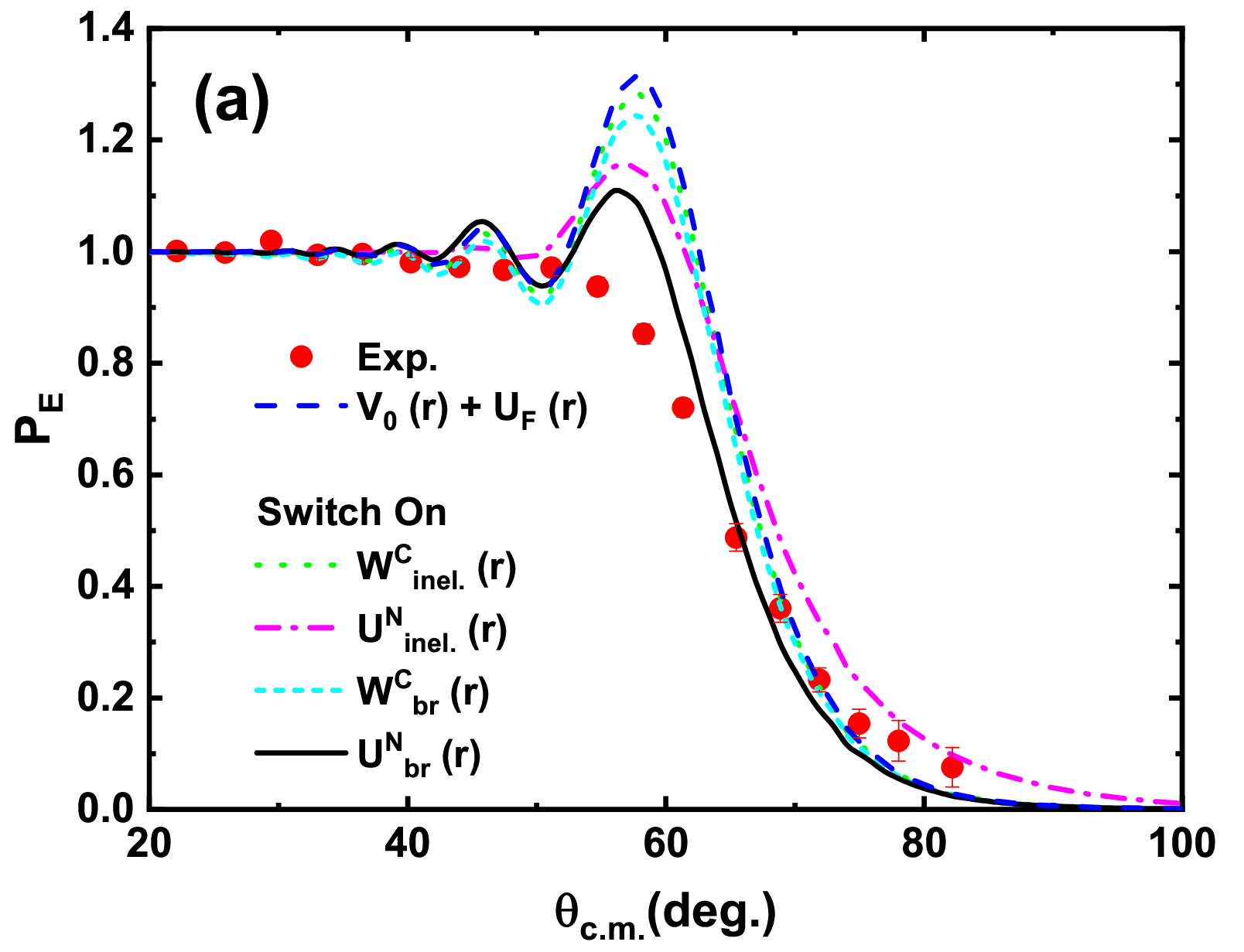}
\includegraphics[width=0.495\linewidth]{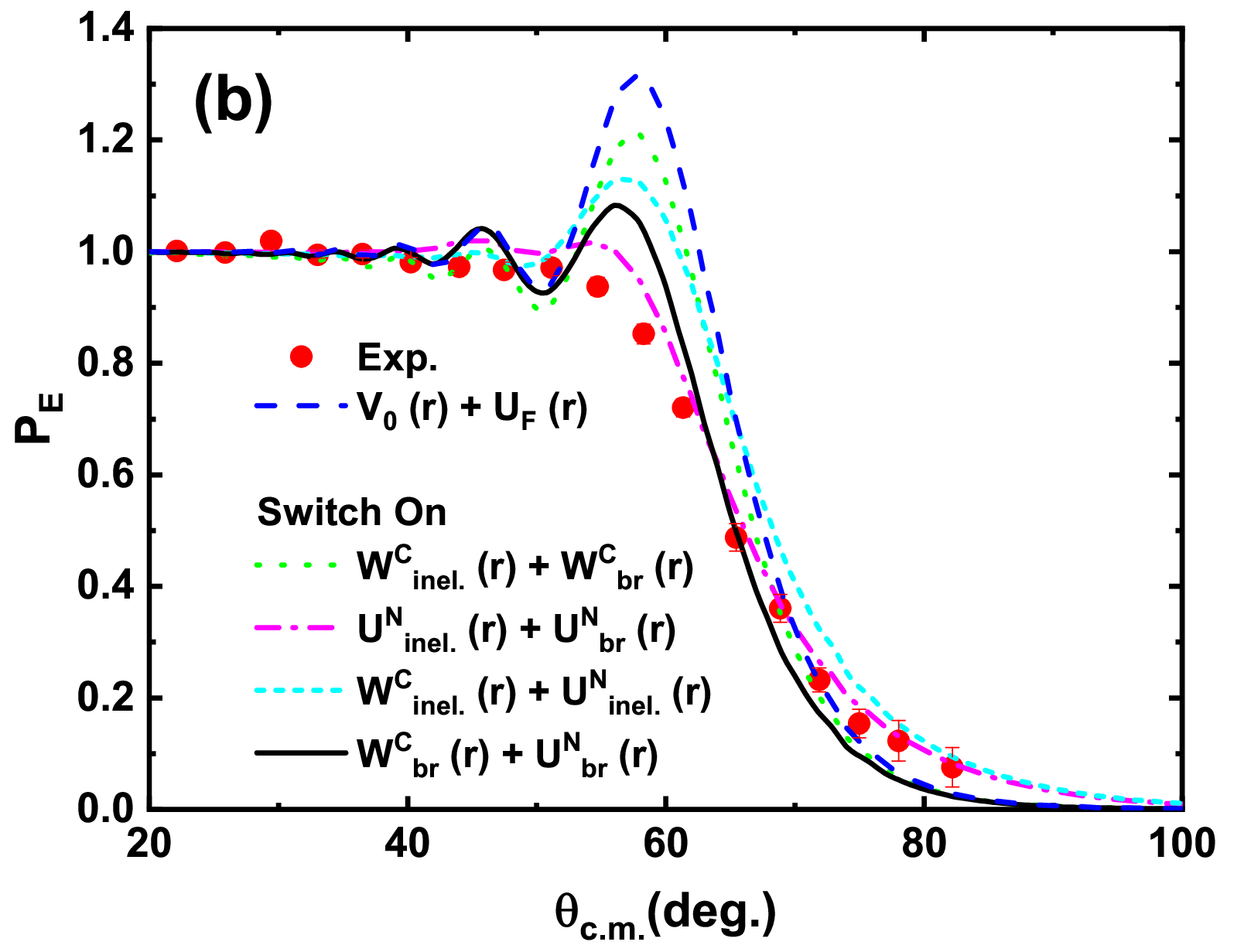}
\end{tabular}
\caption{Similar to Fig.~\ref{om_cal}, the graphs show the contribution of each channel. $P_{\textrm{el}}$ ratios obtained from the (a) single channel and (b) two channel coupling for the $^{17}$Ne + $^{208}$Pb system. The dashed blue lines denote the theoretical $P_{\textrm{el}}$ ratios obtained from $V_{0} (r)$ + $U_{\textrm{F}} (r)$ potential. In the left panel, the dotted green, dash-dotted magenta, short dashed cyan, and solid black lines represent the $P_{\textrm{el}}$ ratios when the potential corresponding to each single channel is switched on at $V_{0} (r)$ + $U_{\textrm{F}} (r)$ potential. Conversely, in the right panel, the dotted green, dash-dotted magenta, short dashed cyan, and solid black lines represent the $P_{\textrm{el}}$ ratios when the potential corresponding to two channel coupling is switched on at $V_{0} (r)$ + $U_{\textrm{F}} (r)$ potential.}
\label{om_channel}
\end{figure}

According to Ref.~\cite{ovej23}, the authors remarked that the Coulomb excitation interaction due to two low-lying E2 resonance states was not the main contributor to the imaginary part of the optical model potential. Therefore, in this study, we examined the contribution of each channel and determined the main contributor. To this end, we examined the change in the elastic scattering cross section by adding (or switching on) each channel to $V_{0} (r)$ + $U_{\textrm{F}} (r)$ potential and investigate the extent to which the contribution of each channel affects the rainbow peak. We also investigated the contributions of paired channels (two-channel coupling).

Fig.~\ref{om_channel} (a) clearly shows that the $P_{\textrm{el}}$ ratio due to the nuclear part of the breakup reaction potential ( solid black line) has the largest suppression. This can also be confirmed by the values of the reaction cross sections listed in Table~\ref{cross-sec}. However, it is unclear whether the contribution of a specific channel causes the rainbow peak to disappear. To clearly observe the contributions of specific channels to the disappearance of the rainbow peak, we paired (or coupled) each channel with its Coulomb, nuclear, inelastic, and breakup reaction components, respectively. Consequently, we obtained Fig.~\ref{om_channel} (b), from which we can carefully infer that the largest contribution to the suppression of the rainbow peak is due to the nuclear component[switch on $U^{\textrm{N}}_{\textrm{inel.}} (r)$ + $U^{\textrm{N}}_{\textrm{br}} (r)$; the dash-dotted magenta line]. The contribution of the breakup reaction [switch on $W^{\textrm{C}}_{\textrm{br}} (r)$ + $U^{\textrm{N}}_{\textrm{br}} (r)$; the solid black line] are also larger than that of inelastic scattering [switch on $W^{\textrm{C}}_{\textrm{inel.}} (r)$ + $U^{\textrm{N}}_{\textrm{inel.}} (r)$; short dashed cyan line]. Finally, the contributions of Coulomb excitation interaction due to two low-lying E2 resonance states [switch on $W^{\textrm{C}}_{\textrm{inel.}} (r)$ + $U^{\textrm{C}}_{\textrm{br}} (r)$; dotted green line] are relatively small. Thus, we can confirm that this contribution is not the primary one, as mentioned above~\cite{ovej23}. Consequently, it can be interpreted that the $^{17}$Ne projectile can be more easily excited or broken up by nuclear interaction than by the Coulomb interaction generated by the target nucleus. Including all the channels, we can obtain significantly better results, as shown in Fig. ~{\ref{om_cal}} (a).

\subsection{CC calculations}
Thus far, we have calculated the elastic scattering, inelastic scattering, breakup reaction, and fusion cross sections for the $^{17}$Ne + $^{208}$Pb system using the OM. In Ref.~\cite{ovej23}, the $^{15}$O production cross section was calculated by treating $\frac{5}{2}^{-}$ state [$E_{x}$ = 1.764 MeV] as the breakup channel using the CC method.
Here, we attempt to recalculate the same system using the CC method~\cite{hagi99} for comparison with the analysis in Ref.~\cite{ovej23}. We examined how the coupling effects of the potentials generated in the OM were realized within the CC framework. To examine the effect of breakup within CC in the CC calculation, the DPP potentials associated with each coupling were excluded from the OM potential. Moreover, the breakup potential $U_{\textrm{br}}$ was replaced with its counterpart in the CC method. It is to confirm the coupling effect of the two resonance states, the $\frac{5}{2}^{-}$ and the $\frac{3}{2}^{-}$ state, in the CC framework, through the potential obtained in our extended optical model.

First, we perform the CC calculations with set (A) of Table~\ref{parameters_17ne}, which has no nuclear components for the inelastic channel, and the $\frac{5}{2}^{-}$ state [$E_{x}$ = 1.764 MeV] and $\frac{3}{2}^{-}$ state [$E_{x}$ = 2.692 MeV] are treated as the breakup interactions. Note that $\frac{3}{2}^{-}$ state [$E_{x}$ = 2.692 MeV] is added to the CC in Ref.~\cite{ovej23}.
For reference, the nuclear deformation length ($\delta_{n}$) corresponding to the $\frac{5}{2}^{-}$ and $\frac{3}{2}^{-}$ states are $\delta_{1}$ = 2.14 and $\delta_{2}$ = 1.82, respectively.

\begin{figure}
\begin{tabular}{c}
\includegraphics[width=0.495\linewidth]{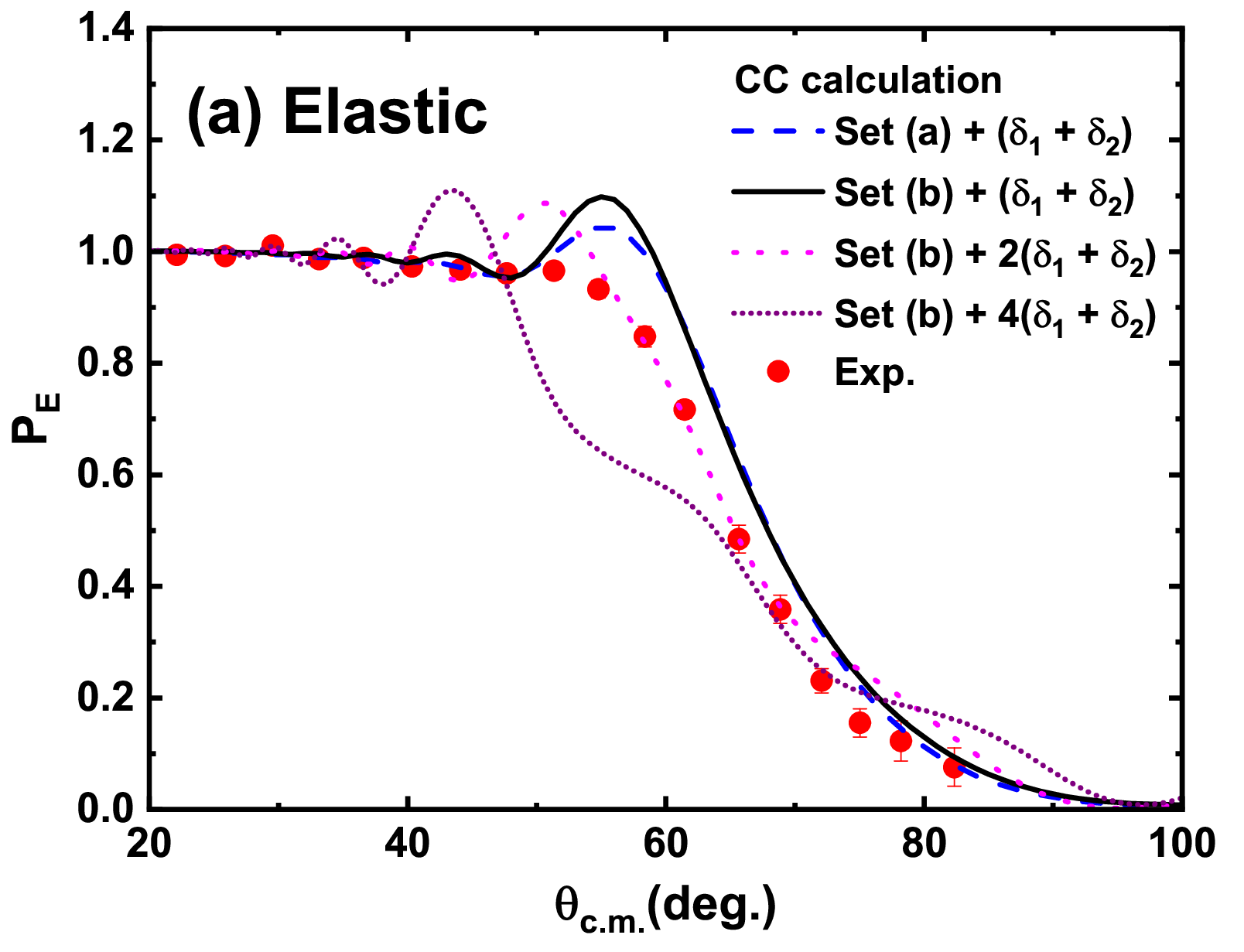}
\includegraphics[width=0.495\linewidth]{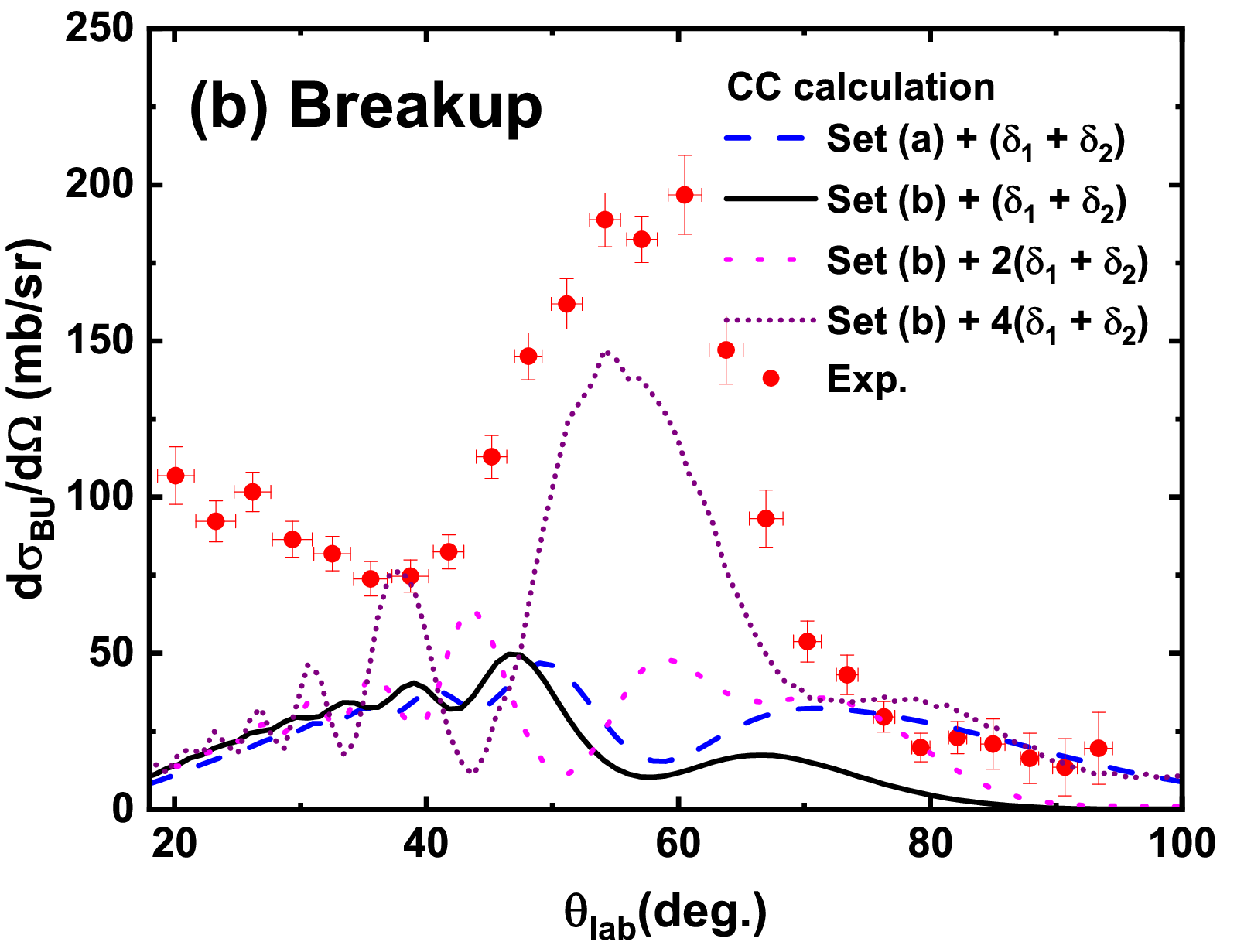}
\end{tabular}
\caption{Similar to Fig.~\ref{om_cal}, the graphs show the results of the CC calculations. The dashed blue and solid black lines represent the theoretical angular distribution of the breakup cross sections obtained from the CC calculations using sets (A) and (B) with two nuclear deformation lengths ($\delta_{1}$ + $\delta_{2}$) in Table~\ref{parameters_17ne}, respectively. The dotted magenta and short-dotted violet lines represent those obtained from the CC calculations using the set (B) with double and quadruple of two nuclear deformation lengths, respectively.}
\label{cc_cal}
\end{figure}

In Fig.~{\ref{cc_cal}}, we plotted the theoretical $P_{\textrm{el}}$ ratio and angular distribution of the breakup cross section by the set (A) parameters as the dashed blue line. Because the theoretical angular distribution of the breakup reaction cross section is not significantly large, as shown in Fig.~{\ref{cc_cal}} (b), the theoretical $P_{\textrm{el}}$ ratio does not sufficiently suppress the experimental $P_{\textrm{el}}$ data.

Next, we perform the CC calculations using set (B) of Table~\ref{parameters_17ne} which considers the additional coupling of the nuclear components for the inelastic channel and the $\frac{5}{2}^{-}$ and $\frac{3}{2}^{-}$ states for the breakup interaction. The theoretical $P_{\textrm{el}}$ ratio and angular distribution of the breakup cross section are plotted as a solid black line in Fig.~{\ref{cc_cal}}. However, the figure shows that the change in the theoretical $P_{\textrm{el}}$ ratios is also insignificant because there is almost no change in the breakup cross section, similar to the calculation of set (A).
This means that, aside from the two low-lying states discussed above, additional channels, such as higher continuum states or transfers, are necessary to describe the breakup cross section. Therefore, in our OM calculations, we inserted a differential form of the Woos-Saxon potential, corresponding to the contribution of the nuclear component of the breakup reaction, to consider the additional channels.

In the CC formalism, considering all channels contributing to the breakup reaction can be challenging. Thus, we simply adjust the nuclear deformation length mentioned above to determine how the theoretical $P_{\textrm{el}}$ ratios and angular distribution of the breakup cross section change, thereby testing the contribution of channels related to nuclear interaction. The contributions of the Coulomb interactions are presented at approximately 40$^{\circ}$ and show results similar to those of the CC analysis in Ref.~\cite{ovej23}.

Furthermore, the results were examined by changing the nuclear deformation length two to four times. The results are plotted as dotted magenta, short-dashed green, and short-dotted violet lines in Fig. ~\ref{cc_cal}. As the nuclear deformation strength increases, the breakup reaction cross section increases and gradually approaches that of the experimental data. This implies that the channels related to nuclear interaction are crucial in reproducing the experimental data of $P_{\textrm{el}}$ ratios and angular distribution of the breakup cross section.

According to our CC calculations, although an additional excited state [$\frac{3}{2}^{-}$ state ($E_{x}$ = 2.692 MeV)] is added to the calculations performed in Ref.~\cite{ovej23}, the result could not explain the $^{15}$O production cross section.

\section{Conclusions}
\label{con}
In this study, we investigated the elastic scattering, inelastic scattering, breakup reaction, and fusion reaction of the $^{17}$Ne + $^{208}$Pb system using the OM and CC methods. Although only the experimental data of the elastic scattering and breakup cross sections were available for the $^{17}$Ne + $^{208}$Pb system, we successfully performed simultaneous $\chi^{2}$ analysis using the theoretical total fusion cross section obtained from the coupled channel method.

In addition, we investigated the contribution of each channel and determined its primary contribution. As a result, we can infer that the dominant contribution to the suppression of the rainbow peak was from the nuclear component, and the contributions of the Coulomb excitation interaction by the two low-lying E2 resonance states were relatively small.

From the simultaneous $\chi^{2}$ analysis of the $^{17}$Ne + $^{208}$Pb system, we can infer a strong suppression effect in the elastic scattering cross section because of the nuclear interaction between the projectile and target nucleus, rather than the Coulomb interaction, as observed in neutron-rich nuclei. The contribution of the direct reaction, comprising the inelastic scattering and breakup reaction cross-sections, accounted for almost half of the total reaction.

Finally, we performed the CC calculation using the parameters obtained from our OM calculation, which considered an additional excited state [$\frac{3}{2}^{-}$ state ($E_{x}$ = 2.692 MeV)] compared with the calculations performed in Ref.~\cite{ovej23}; however, the additional excited state also could not explain properly the $^{15}$O production cross section. Nevertheless, based on the changes in the deformation relevant to nuclear interactions, we confirmed that the suppression of the rainbow peak was mainly due to nuclear interactions.

It is well known that a large part of the elastic breakup owing to long-distance Coulomb interactions between the target and projectile can be explained using methods such as continuum-discretized-coupled channels (CDCC) and DPP using $dB(E1)/dE$~\cite{andr94}. However, direct reactions caused by nuclear interactions that occur at short distances, in which energy is easily exchanged, have much more complex reaction mechanisms. In particular, nonelastic breakup channel may include a various effects, such as the transfer or stripping of each fragment to the target after the projectile breaks up. In the present study, we effectively divided these complex reaction channels into major ones using effective DPPs.

\section*{Acknowledgment}
\label{sec:orgae7f2b2}
This work was supported by the National Research Foundation of Korea
(Grant Nos. NRF-RS-2023-00276361, NRF-2021R1F1A1051935, NRF-2021R1F1A1046575, NRF-2021R1A6A1A03046957, NRF-2020R1A2C3006177, NRF-2017R1E1A1A01074023 and NRF-2013M7A1A1075764) and MSIT (No. 2018R1A5A1025563).

\end{document}